\documentclass{article}
\usepackage{spconf,amsmath,graphicx}

\usepackage{multirow} 
\usepackage{cite} 
\usepackage{makecell} 
\usepackage{algorithm} 
\usepackage{algorithmic} 

\title{EDITSPEECH: A TEXT BASED SPEECH EDITING SYSTEM \\USING PARTIAL INFERENCE AND BIDIRECTIONAL FUSION}
\name{Daxin Tan$^1$, Liqun Deng$^2$, Yu Ting Yeung$^2$, Xin Jiang$^2$, Xiao Chen$^2$, Tan Lee$^1$}
\address{
  $^1$Department of Electronic Engineering, The Chinese University of Hong Kong, Hong Kong\\
  $^2$Huawei Noah’s Ark Lab, Shenzhen, China}

\begin{document}

\maketitle

\begin{abstract}
This paper presents the design, implementation and evaluation of a speech editing system, named EditSpeech, which allows a user to perform deletion, insertion and replacement of words in a given speech utterance, without causing audible degradation in speech quality and naturalness. The EditSpeech system is developed upon a neural text-to-speech (NTTS) synthesis framework. Partial inference and bidirectional fusion are proposed to effectively incorporate the contextual information related to the edited region and achieve smooth transition at both left and right boundaries. Distortion introduced to the unmodified parts of the utterance is alleviated. The EditSpeech system is developed and evaluated on English and Chinese in multi-speaker scenarios. Objective and subjective evaluation demonstrate that EditSpeech outperforms a few baseline systems in terms of low spectral distortion and preferred speech quality. Audio samples are available online for demonstration\footnote{https://daxintan-cuhk.github.io/EditSpeech/}.
\end{abstract}

\begin{keywords}
speech editing, speech synthesis, prosody
\end{keywords}

\section{Introduction}

Nowadays audio sharing via social media has become a prevalent activity in our daily life. With mobile apps like Himalaya and Instagram, users can conveniently record their own speech and share with others. When making a long speech recording, e.g., telling a story, describing procedures, unintentional mistakes of speaking like mispronunciations, missing words, stuttering, etc., are inevitable, especially for non-professional speakers. Even if the mistakes only affect locally a small part of the audio, the user may need to re-do the whole recording from the beginning, in order to maintain a coherent speech quality and speaking style. It would be highly desirable to allow the user to edit the recorded speech, e.g., insert missed words, replace mispronounced words, and/or remove unwanted speech or non-speech events, without degrading the quality and naturalness of the edited speech. This demand motivates our present study of designing a novel speech editing system, named EditSpeech.

\begin{figure}[h]
  \centering
  \includegraphics[width=\linewidth, trim=0 20 0 0]{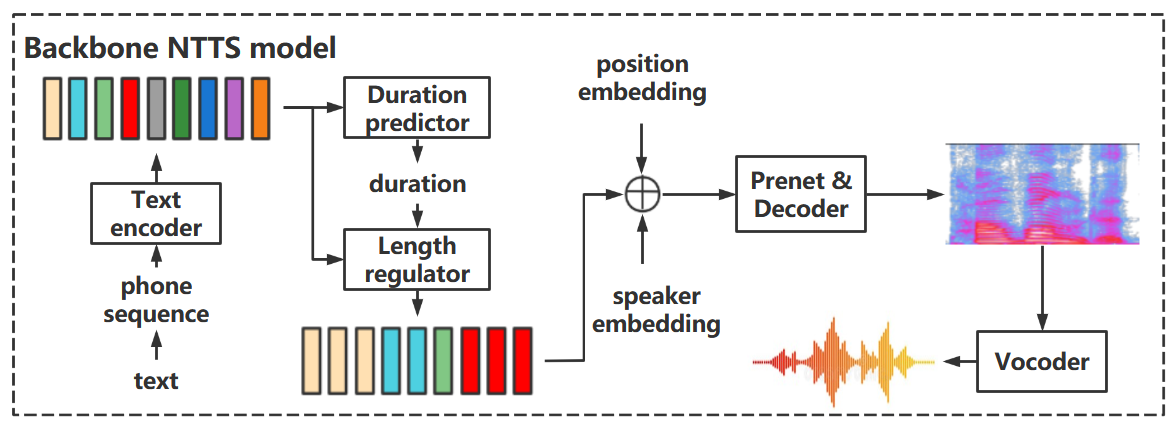}
  \caption{Backbone NTTS model}
  \label{fig:backbone_TTS_model}
  \vspace{-1em}
\end{figure}

\begin{figure*}[t]
  \centering
  \includegraphics[width=\linewidth, trim=0 10 0 0]{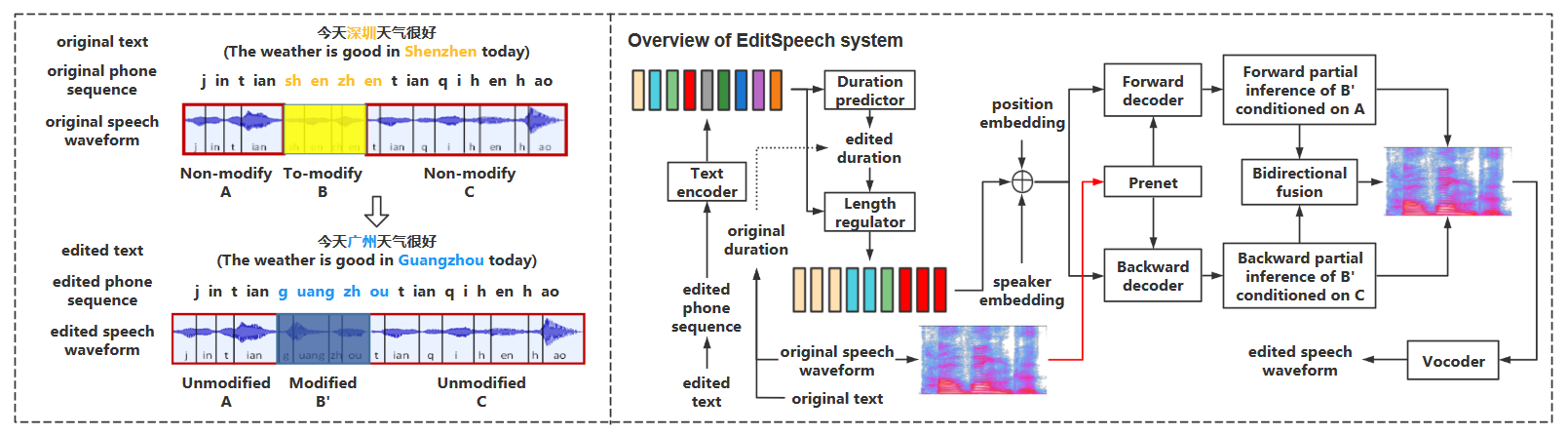}
  \caption{Overview of EditSpeech system and a Chinese speech editing example (replacement)}
  \label{fig:overview_EditSpeech_system}
  \vspace{-1em}
\end{figure*}

There were a few previous attempts to developing speech editing systems \cite{whittaker2004semantic, rubin2013content, baume2018contextual, jin2017voco, morrison2021context,descript}. The VOCO system \cite{jin2017voco} was developed for English speech editing in a multi-speaker scenario. Speech signals containing the words for inserting or replacing are generated by unit-selection speech synthesis. As no contextual information is taken into consideration, speech prosody near boundaries of the edited regions could be non-smooth and unnatural. In \cite{morrison2021context}, context-aware prosody correction was applied in single-speaker English speech editing. The speech segments to be inserted were retrieved from other utterances of the same speaker. Target duration and pitch parameters were predicted from the context, and then prosodic modification was realized by applying the TD-PSOLA algorithm \cite{moulines1990pitch}, followed by de-noising and de-reverberation \cite{su2020hifi}. An obvious limitation of this system is that the words to insert or replace may not be found in the available speech data of the same speaker.

In the present study, we tackle the problem of speech editing on the basis of neural text-to-speech synthesis (NTTS) \cite{shen2018natural,ping2018deep,yu2019durian,ren2020fastspeech}. An NTTS based speech editing system, named EditSpeech, is developed and evaluated in this study. The key idea of EditSpeech is that, given a speech utterance, we divide it into ``to-modify'' and ``non-modify'' regions according to the edited text and speech-text alignment, and generate the new ``modified'' speech frames using NTTS conditioned on the ``non-modify'' frames. Several elaborate designs are proposed. First, partial inference is adopted in the speech generation process, i.e., only the ``to-modify'' region are calculated during inference to produce ``modifed'' frames, while the frames in ``non-modify'' region are directly copied to produce ``unmodified'' frames to minimize unnecessary distortion. Second, a duration based auto-regressive (AR) NTTS model is employed for generating ``modified'' frames, and the decoding logic is implemented in both forward and backward directions to maximize the use of the left and right ``non-modify'' frames as contextual condition respectively. Third, a bidirectional fusion process is followed to select the best generated frames from NTTS model. In this way, contextual information related to the edited region is integrally utilized and smooth transition at the boundaries can be achieved. 

The EditSpeech system is more efficient than unit-selection systems like VOCO, as search and selection of candidate units are not required in usage. EditSpeech can support arbitrary change of text content and does not require additional recording from the speakers concerned. The current version of EditSpeech is developed for both English and Chinese in the multi-speaker scenario.

\vspace{-1em}
\section{Related work}

Currently, the most widely used NTTS models can be categorized into two types: attention based AR models and duration based non-autoregressive (NAR) models. Tacotron2 \cite{shen2018natural} and Transformer TTS \cite{li2019neural} are the typical examples of the former, in which the text embedding and acoustic feature are aligned with a location-sensitive attention or multi-head self-attention, and the decoding of the current time step always conditions on the result of its previous step. In contrast, FastSpeech2 \cite{ren2020fastspeech} and Glow-TTS \cite{kim2020glow} are the representatives of the latter. They employ an extra duration predictor to address the alignment between text and acoustic frames, thus the decoding of all the time steps are conducted in parallel without internal dependency. However, these models cannot be directly employed to solve speech editing problem, as the attention based AR ones are not able to control the duration of the generated ``modified'' frames, while the duration based NAR ones fail to utilize the neighboring speech context due to the parallel generation. EditSpeech adopts a duration based AR model as the backbone model, which is shown in Figure \ref{fig:backbone_TTS_model}, to take both the advantages of these two types of models. The backbone model is similar to that in DurIAN \cite{yu2019durian} and Patnet \cite{wang2021patnet}, but our scheme differs in that we refine the predicted duration and use the partial inference for speech generation in the editing scenario. 

On the other hand, to make the edited speech as much natural as possible, EditSpeech adopts two decoders to produce acoustic frames, one in the left-to-right direction and the other in the right-to-left direction. This is relatively novel in speech generation domain. The most similar scheme may be the one in \cite{zheng2019forward}. They adopt two unidirectional decoders and maximize the agreement between forward and backward decoding sequences as regularization for the training of TTS system. Their target is to alleviate the ``exposure bias'' problem, and they only use one decoder in the synthesis stage. Our work is different in that we aim to better utilize the context, and we use both decoders in the editing scenario. Moreover, similar scheme can be seen in neural machine translation field in \cite{zhou2019synchronous}, which aims to utilize both the historical and future information in the text-text generation. Different from them, we focus on the context utilization on both the text and speech side. Our decoders not only take in the encoder output as the text-side guidance, but also generate speech frames conditioned on the contextual acoustic frames from ``non-modify'' regions.

\vspace{-1em}
\section{The EditSpeech System}
\vspace{-0.5em}
Figure \ref{fig:overview_EditSpeech_system} gives an overview of the EditSpeech system, where speech editing is exemplified through a replacement operation of a Chinese utterance. The edited text and the original speaker identity are used to derive the hidden representation. Duration information obtained from the original text and original speech are used as reference for the duration of edited speech. Based on the original speech's mel-spectrogram and the hidden representation, the two decoders predict mel-spectrogram in a partial inference manner. The two predicted mel-spectrograms are then fused into a single edited mel-spectrogram, which is converted by the vocoder into edited speech waveform.

The EditSpeech system comprises an acoustic model, a duration predictor and a vocoder. The acoustic model is made up of a text encoder, a speaker encoder, a length regulator, a prenet, a forward decoder and a backward decoder. 

\begin{figure}[t]
  \centering
  \includegraphics[width=\linewidth, trim=0 10 0 0]{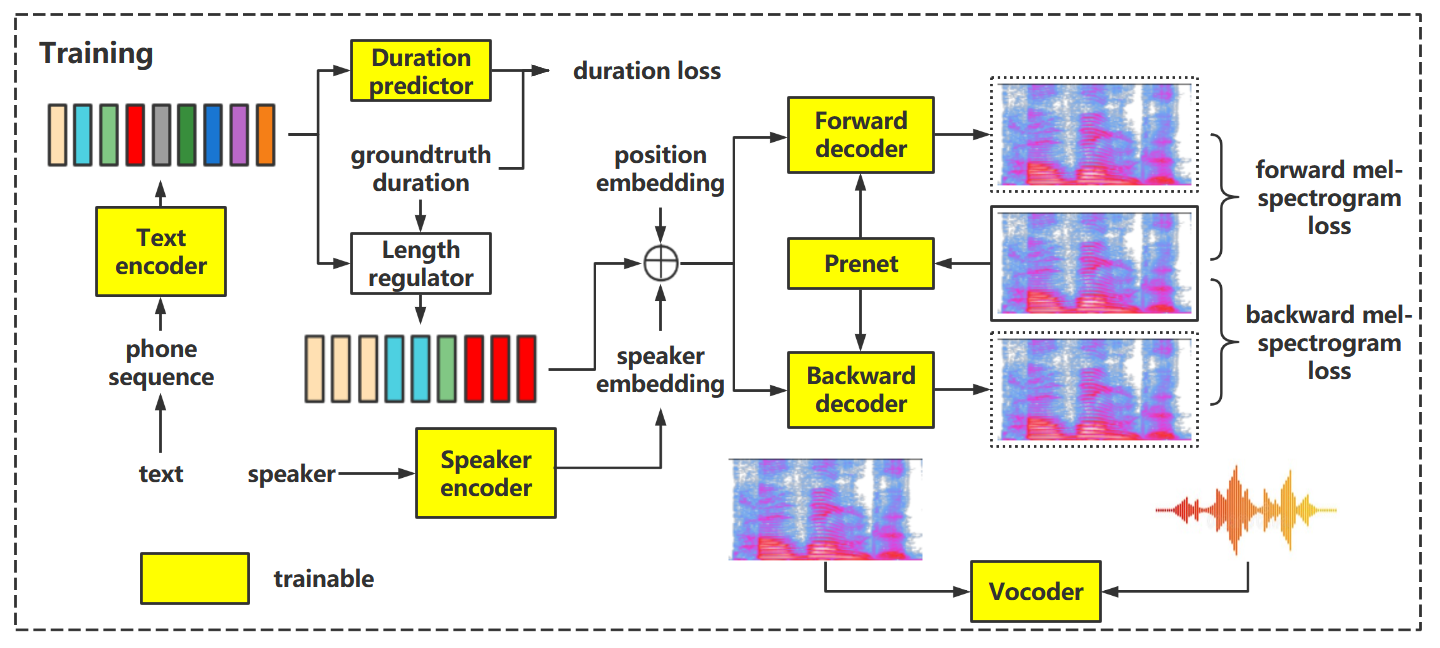}
  \caption{Training process of EditSpeech system}
  \label{fig:training_EditSpeech_system}
  \vspace{-1em}
\end{figure}

The training process for the EditSpeech system is shown in Figure \ref{fig:training_EditSpeech_system}. Speech utterances with text transcription and speaker identity are used for training. The texts are converted into phone sequences using a grapheme-to-phoneme (G2P) module. Mel-spectrogram is computed from the speech utterances using the same signal processing configuration as in the Tacotron2 model \cite{shen2018natural}. The ground-truth phone duration is obtained by an HMM based forced aligner. 

The phone sequence is first processed by the text encoder to derive the phone-level text embedding. The speaker encoder generates an utterance-level speaker embedding from the speaker identity. Frame-level position embeddings are generated as interpolated values from 0 to 1, which represent the relative position of individual frames within a phone. The text embedding is expanded into frame-level embedding according to the ground-truth phone duration by the length regulator. It is then concatenated with the position embedding and the speaker embedding to construct frame-level hidden representation.

A forward decoder and a backward decoder are used to synthesize mel-spectrograms in the left-to-right and the right-to-left directions respectively. These two decoders share a common prenet and a common linear layer. Each decoder comprises two unidirectional LSTM. For the left-to-right synthesis of speech frame at time $t$, the ground-truth mel-spectrum of preceding frame $m_{t-1}$ is processed first by the prenet. The prenet output is concatenated with the frame-level hidden representation of the preceding frame, denoted as $h_{t-1}$, and fed into a unidirectional LSTM to derive the context vector. Then the context vector and $h_{t}$ are taken by another unidirectional LSTM and a linear layer to predict a mel-spectrum representing the current frame, denoted as $\overrightarrow{m}_t$. The mean squared error (MSE) between $\overrightarrow{m}_t$ and the ground-truth mel-spectrum $m_t$ is used for the computation of the forward mel-spectogram loss $L_{\overrightarrow{Mel}}$. The prediction of mel-spectrogram in the right-to-left direction is conducted in a similar manner except that $m_{t+1}$ and $h_{t+1}$ are involved to generate $\overleftarrow{m}_t$. Let the backward mel-spectrogram loss be denoted by $L_{\overleftarrow{Mel}}$. The text encoder, the speaker encoder, the prenet, the forward decoder and the backward decoder are jointly trained to minimize $L_{\overrightarrow{Mel}}+L_{\overleftarrow{Mel}}$. 

The duration predictor aims to predict phone-level duration from the phone sequence. It is trained by minimizing the duration loss, which is the MSE between the logarithms of ground-truth and predicted duration. 

There are many different designs and implementations of vocoder for converting mel-spectrogram into speech waveform \cite{valin2019lpcnet,prenger2019waveglow,yamamoto2020parallel,kumar2019melgan,kong2020hifi}. In our work, the HiFi-GAN vocoder \cite{kong2020hifi} is adopted in consideration of its good quality and computational efficiency. The vocoder is trained from a open-sourced pre-trained version ``UNIVERSAL\_V1'' \footnote{https://github.com/jik876/hifi-gan}.

\vspace{-1em}
\section{Speech Editing Operations}

With a properly trained EditSpeech system, a user can perform speech editing by carrying out one of the three operations, namely deletion, insertion and replacement.

\vspace{-1em}
\subsection{Deletion}
The deletion operation allows the user to remove a section of speech waveform that corresponds to certain specified words. By forced alignment, the system locates the start and end time of the phones to be deleted. The corresponding speech frames in the input speech's mel-spectrogram are removed to obtain the edited mel-spectrogram, which is converted into speech waveform by the trained vocoder. 

\vspace{-1em}
\subsection{Insertion and replacement}

Speech editing involving insertion and/or replacement of words is more complex and has higher requirement than deletion as the edited speech contains newly created content. In the operation of inserting words, the user needs to specify: 1) word position for the insertion; and 2) the text to insert. For replacing words, the users should specify: 1) the first and the last words to be replaced. 2) the new text.

The implementation of insertion and replacement is demonstrated through an example. Readers may refer to Figure \ref{fig:overview_EditSpeech_system}. Without loss of generality, the mel-spectrogram of the original speech is divided into three parts, denoted as $[m_A, m_B, m_C]$, which correspond to three parts of text $[T_A, T_B, T_C]$. For $i\in\{A,B,C\}$, $m_i$ contains a sequence of frame-level mel-spectra and $T_i$ is a sequence of words. Our goal is to obtain an edited speech waveform with the text content changed to $[T_A, T_{B'}, T_C]$, where $T_{B'}$ is the text content to replace $T_B$. Insertion is considered as the special case where $m_B=T_B=\emptyset$.

\vspace{-1em}
\subsubsection{Preparation of phone sequence}
With the G2P module, the original text and the replacement text are converted to phone sequence, denoted as $[P_A, P_B, P_C]$ and $P_{B'}$ respectively. $P_B$ in the original phone sequence is changed to $P_{B'}$ to give the edited phone sequence $[P_A, P_{B'}, P_C]$.

\subsubsection{Duration sequence prediction and refinement}
The original duration sequence is denoted by $[dur_A, dur_B, \\dur_C]$, where  $dur_i, i\in\{A,B,C\}$ is a sequence of phone duration in frame. They are obtained by forced alignment process with the original text, The predicted duration sequence $[dur_A^p, dur_{B'}^p, dur_C^p]$ is obtained from the duration predictor with the edited phone sequence $[P_A, P_{B'}, P_C]$ as input.

To ensure that the speaking rate is consistent in the modified region ($B'$) and unmodified regions ($A$, $C$) of speech, the predicted duration $dur_{B'}^p$ is refined by referring to the original and predicted duration of the unmodified region, i.e.,
$$dur_{B'}=dur_{B'}^p*(\sum dur_A+\sum dur_C)/(\sum dur_A^p+\sum dur_C^p)$$

$dur_B$ of the original duration sequence is replaced by $dur_{B'}$ to construct the edited duration sequence $[dur_A, dur_{B'}, \\dur_C]$, and $t_{tot}=\sum dur_A+\sum dur_{B'}+\sum dur_C$ is the total duration of edited speech in frame.

\subsubsection{Partial inference}

The edited phone sequence, the edited duration sequence, the speaker embedding and the position embedding all serve as the inputs. The frame-level hidden representation $h$ are derived from these inputs by the text encoder, speaker encoder and length regulator, which is then used for mel-spectrogram generation in both forward and backward direction.

\begin{figure}[t]
  \centering
  \includegraphics[width=\linewidth, trim=20 50 60 40]{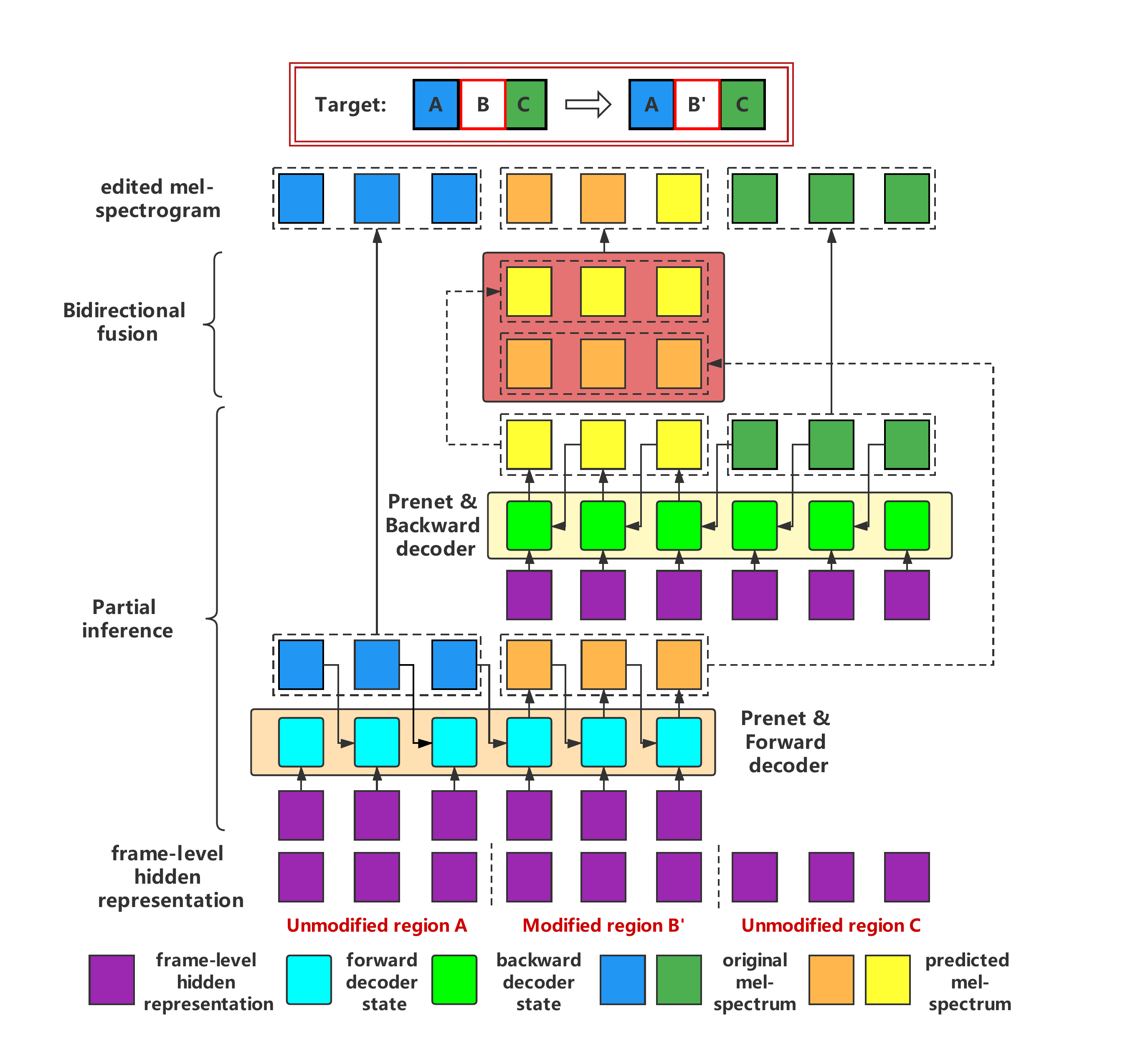}
  \caption{Details of partial inference and bidirectional fusion}
  \label{fig:partial_inference_and_bidirectional_fusion}
  \vspace{-1em}
\end{figure}

Different from the fully auto-regressive generation in normal NTTS system, our system generates the mel-spectrogram in the partial inference manner for both forward and backward direction, of which the detail is shown in Figure \ref{fig:partial_inference_and_bidirectional_fusion}. Specifically, for each time step in the unmodified region, the predicted frame is discarded, and the original frame is fed to the prenet and recurrent decoder for the prediction of next frame. In the contrary, for each time step in the modified region, the predicted frame is fed to the prenet and recurrent decoder for the prediction of next frame.

(i) The partial inference process in forward direction:

\textbf{Initialization:} $m_0=\mathbf{0}$, $h_0=\mathbf{0}$

\textbf{The unmodified region:}

for $t=1$ to $t=\sum dur_A$
\begin{eqnarray}    
\label{eq:1}
\qquad\qquad \overrightarrow{m}_t=\overrightarrow{Decoder}(Prenet(m_{t-1}), h_{t-1},  h_t)\nonumber
\end{eqnarray}

\textbf{The modified region:}

for $t=\sum dur_A+1$ to $t=\sum dur_A+\sum dur_{B'}$
\begin{eqnarray}  
\qquad\qquad \overrightarrow{m}_t=\overrightarrow{Decoder}(Prenet(\overrightarrow{m}_{t-1}), h_{t-1},  h_t)\nonumber
\end{eqnarray}

(ii) The partial inference process in backward direction:

\textbf{Initialization:} $m_{t_{tot}+1}=\mathbf{0}$, $h_{t_{tot}+1}=\mathbf{0}$

\textbf{The unmodified region:}

for $t=t_{tot}$ to $t=\sum dur_A+\sum dur_{B'}+1$
\begin{eqnarray}    
\label{eq:2}
\qquad\qquad \overleftarrow{m}_t=\overleftarrow{Decoder}(Prenet(m_{t+1}), h_{t+1},  h_t)\nonumber
\end{eqnarray}

\textbf{The modified region:}

for $t=\sum dur_A+\sum dur_{B'}$ to $t=\sum dur_A+1$
\begin{eqnarray}  
\qquad\qquad \overleftarrow{m}_t=\overleftarrow{Decoder}(Prenet(\overleftarrow{m}_{t+1}), h_{t+1},  h_t)\nonumber
\end{eqnarray}

\vspace{-1em}
\subsubsection{Bidirectional fusion}

The partial inference process predicts the mel-spectrogram based on the frames that they have encountered. As a result, forward decoder and backward decoder guarantee the fluency at the left and the right boundaries of edit region respectively. To improve the fluency and naturalness at both boundaries, the predicted mel-spectrogram from both forward decoder and backward decoder ($\overrightarrow{m}_t$ and $\overleftarrow{m}_t$) are fused. This process is named as bidirectional fusion and the detail is also shown in Figure \ref{fig:partial_inference_and_bidirectional_fusion}. It should be noted that, $\overrightarrow{m}_t$ and $\overleftarrow{m}_t$ are frame-synchronous. In the modified region, the frame-level $L2$-norm differences between two predicted mel-spectrogram are first calculated, then the frame with the least $L2$-norm difference is selected as the fusion point. 
$$t_{fusion}=\mathop{\arg\min}_{\sum dur_A<t\leq\sum (dur_A+dur_{B'})} ||\overrightarrow{m}_t-\overleftarrow{m}_t||_2$$

For each frame in the unmodified region, the original mel-spectrum $m_t$ directly serves as the edited mel-spectrum $m_t^e$. In the modified region, for each frame before the fusion point, the forward predicted mel-spectrum $\overrightarrow{m}_t$ is selected as the edited mel-spectrum $m_t^e$. For each frame after the fusion point, the backward predicted mel-spectrum $\overleftarrow{m}_t$ is selected as the edited mel-spectrum $m_t^e$. The edited mel-spectum $m_t^e$ are merged to construct the edited mel-spectrogram, which is then fed to the trained vocoder to generate the waveform.

$$ m_t^e=\left\{
\begin{array}{rcl}
m_t, & & t\in[1, \sum dur_A]\\
\overrightarrow{m}_t, & &t\in(\sum dur_A, t_{fusion}] \\
\overleftarrow{m}_t, & &t\in(t_{fusion}, \sum dur_A+\sum dur_{B'}]\\
m_t, & &t\in(\sum dur_A+\sum dur_{B'}, t_{tot}]
\end{array}
\right.$$

\section{Experiment}

\subsection{Data}

Four speech editing systems are developed on four datasets \cite{ljspeech17,yamagishi2019cstr,shi2020aishell} respectively. The overview of datasets is shown in Table \ref{tab:dataset}. For each speaker, we select 99\% of speech utterances as the training set, and the remaining 1\% as the test set for speech editing.
\vspace{-1em}

\begin{table}[h]
\caption{Datasets for system development}
\label{tab:dataset}
\centering
\begin{small}
\begin{tabular}{c|c|c|c}
\hline
Dataset & Language & Speakers & Speech quality \\
\hline
LJ-speech & English & 1 female & high\\
\hline
VCTK & English & 109 speakers & high\\
\hline
MST & Chinese & 1 female, 1 male & high\\
\hline
Aishell-3 & Chinese & 218 speakers & moderate\\
\hline
\end{tabular}
\end{small}
\vspace{-1em}
\end{table}

\subsection{Configuration detail}
The audio is sampled to 22050 Hz. The short-time Fourier transform (STFT) is first carried out using 50 ms frame size and 12.5 ms frame hop with the Hann window function. Then the STFT magnitude is converted to the mel-spectrogram based an 80-channel mel filterbank from 0 to 8000 Hz, followed by log dynamic range compression.

Forced alignment is carried out by Montreal Forced Aligner (MFA) \cite{mcauliffe2017montreal}. The G2P modules are different in English and Chinese. For English, the g2pE \cite{g2pE2019} with CMU Pronouncing Dictionary\footnote{http://www.speech.cs.cmu.edu/cgi-bin/cmudict} is adopted to convert English words to phones. For Chinese, a toolkit named pypinyin\footnote{https://github.com/mozillazg/python-pinyin} is first utilized to convert the Chinese characters (Hanzi) to pinyins. Then another pre-trained G2P module in MFA converts the pinyins into global phones. The punctuation like comma is converted to the ``short pause'' symbol in this process. 

The configuration of modules is listed in Table \ref{tab:module}. The Adam optimizer \cite{kingma2014adam} with learning rate of $10^{-3}$ and weight decay of $10^{-6}$ is used in the training process. The batch size is set to be 32. The English/Chinese systems are trained for 100k/200k iterations respectively.

\begin{table}[h]
\caption{Module configuration}
\label{tab:module}
\begin{small}
\centering
\begin{tabular}{l|l}
\hline
Module & Configuration \\
\hline
Text encoder& \makecell[l]{A 512-dim trainble look-up table,\\ 3 Conv1d layers of kernel 5 and channel 512, \\ a 512-dim bidirectional LSTM}\\
\hline
Speaker encoder & A 128-dim trainable look-up table\\
\hline
Prenet & \makecell[l]{Two linear layers of (80, 256) and (256, 256)}\\ 
\hline
Forward decoder & \makecell[l]{Two unidirectional 1024-dim LSTM and \\a shared linear layer of (1024, 80)}\\
\hline
Backward decoder & \makecell[l]{Two unidirectional 1024-dim LSTM and \\a shared linear layer of (1024, 80)}\\
\hline
Duration predictor & \makecell[l]{A two-layer bidirectional 512-dim LSTM \\and a linear layer of (512, 1)}\\
\hline
\end{tabular}
\end{small}
\vspace{-1em}
\end{table}

\subsection{Systems for comparison}
\vspace{-0.5em}
We mainly compare EditSpeech and the baseline systems in terms of insertion and replacement operations. As new content is added in these two operations, the differences between outputs of systems are obvious for evaluation. The baseline systems are listed as below. 

\textbf{Baseline system 1:}
This system is a complete TTS system with the whole edited text and original speaker as input, without considering the original speech. 

\textbf{Baseline system 2:} 
This system is a TTS system with the new text part and original speaker as input, and the original speech is utilized by only simple concatenation. Specifically, the text to insert/replace is used to generate mel-spectrogram, which is then concatenated with the unmodified region of original mel-spectrogram to construct the edited mel-spectrogram.

\textbf{Baseline system 3:} 
This system is a TTS system with the whole edited text and original speaker as input, and the original speech is utilized by frame location and concatenation. Specifically, candidate mel-spectrogram is synthesized as in baseline system 1. Then DTW \cite{muller2007dynamic} of mel-cepstral coefficient (MCEP) is used to align the unmodified region of original and candidate mel-spectrogram. In this way, the modified region of the candidate mel-spectrogram can be located, which is then concatenated with the unmodified region of original mel-spectrogram to construct the edited mel-spectrogram.

\textbf{Baseline system 4:}
This system is similar to the proposed system except that only one left-to-right decoder is adopted and the bidirectional fusion step is removed.

\vspace{-1em}
\section{Result and Discussion}
\vspace{-0.5em}

\begin{table*}[h]
\caption{MCD of reconstructed speech}
\label{tab:reconstruction_MCD}
\centering
\begin{small}
\begin{tabular}{c|c|c|c|c|c|c|c|c|c|c|c|c}
\hline
\multirow{2}{*}{MCD$\downarrow$} &\multicolumn{3}{c|}{LJ-speech} &\multicolumn{3}{c|}{VCTK} &\multicolumn{3}{c|}{MST} &\multicolumn{3}{c}{Aishell-3}\\
\cline{2-13}
~ & Modi. & Unmodi. & Whole & Modi. & Unmodi. & Whole & Modi. & Unmodi. & Whole & Modi. & Unmodi. & Whole\\
\hline
Base.1 &6.20&6.18&6.19 &5.67&4.66&4.88 &5.64&5.51&5.57 &5.93&5.30&5.52\\
\hline
Base.2 &6.46&3.41&4.62 &6.20&2.72&3.53 &6.05&3.70&4.59 &6.59&3.42&4.54\\
\hline
Proposed &\textbf{6.01}&\textbf{3.39}&\textbf{4.43} &\textbf{5.54}&\textbf{2.71}&\textbf{3.38} &\textbf{5.55}&\textbf{3.68}&\textbf{4.34} &\textbf{5.81}&\textbf{3.22}&\textbf{4.08}\\
\hline
\end{tabular}
\end{small}
\vspace{-1em}
\end{table*}

\subsection{Objective evaluation}
\vspace{-0.5em}
In our experiment, 1/3 of the words at the middle location of sentence are masked. The audio part of masked words are first deleted, and then re-synthesized and inserted back into the original location. As the original audio uttered by human is of high naturalness, a lower difference between the edited audio and original audio not only indicates a higher naturalness of edited audio but also demonstrates a better utilization of the text and speech context. Mel-cepstral distortion (MCD) is adopted to measure the difference of edited and original audio, where lower MCD means higher similarity. The MCD evaluation is carried out on three part: the modified region, the unmodified region and the whole utterance. 30 utterances are randomly selected for MCD calculation. The MCD of baseline system 1, 2 and our proposed system are compared, and the results are shown in Table \ref{tab:reconstruction_MCD}.

The results indicate that: 1) In the modified region, the proposed system has the lowest MCD among the three systems, while the baseline system 1 has lower MCD than the baseline system 2. The reason should be that the baseline system 2 synthesize the modified region directly without utilizing both the speech and text context, while the baseline system 1 utilizes the text context but neglects the speech context. In contrast, the proposed system utilizes both the speech and text context. 2) In the unmodified region, the baseline system 2 and the proposed system has much lower MCD then the baseline system 1. As the baseline system 1 synthesizes the speech audio directly from the text without taking the original mel-spectrogram into consideration, the generated mel-spectrogram has large MCD even for the parts that are not intended to modify. For baseline system 2 and the proposed system, the mel-spectrogram of unmodified region is exactly the same as the original one, and the distortion of waveform mainly comes from the vocoder performance. 3) The MCD of whole utterance implicitly reflects the naturalness of the whole utterance. The proposed system exhibits the lowest MCD compared with the baseline system 1 and 2. 

\vspace{-0.5em}
\subsection{Subjective evaluation}
\vspace{-0.5em}
\begin{table}[t]
\caption{MOS of the naturalness and similarity of the edited speech after insertion and replacement}
\label{tab:insert_replace_text_MOS}
\centering
\begin{small}
\begin{tabular}{p{8mm}|p{4.5mm}|p{4.5mm}|p{4.5mm}|p{4.5mm}|p{4.5mm}|p{4.5mm}|p{4.5mm}|p{4.5mm}}
\hline
\multirow{3}{*}{MOS$\uparrow$}  & \multicolumn{4}{c|}{VCTK} & \multicolumn{4}{c}{MST}\\
\cline{2-9}
~ & \multicolumn{2}{c|}{Insert} & \multicolumn{2}{c|}{Replace} & \multicolumn{2}{c|}{Insert} & \multicolumn{2}{c}{Replace} \\
\cline{2-9}
~ & Nat. & Sim.  & Nat. & Sim. & Nat. & Sim.  & Nat. & Sim.\\
\hline
Base.1 &2.97 &2.73 &3.08 &2.91 &2.84 &2.91 &3.01 &3.01\\
\hline
Base.2 &3.13 &2.91 &3.15 &2.93 &2.21 &2.97 &2.22 &3.05\\
\hline
Base.3 &3.14 &2.88 &3.23 &2.88 &3.01 &3.48 &2.95 &3.52\\
\hline
Base.4 &3.10 &2.90 &3.28 &2.93 &3.36 &\textbf{3.70} &3.34 &\textbf{3.77}\\
\hline
Prop. &\textbf{3.34} &\textbf{2.97} &\textbf{3.45} &\textbf{3.03} &\textbf{3.42} &3.54 &\textbf{3.66} &3.69\\
\hline
\end{tabular}
\vspace{-1em}
\end{small}
\end{table}

\begin{table}[t]
\caption{ABX preference of the edited speech after insertion and replacement}
\label{tab:insert_repalce_text_preference}
\centering
\begin{small}
\begin{tabular}{p{8mm}|p{5mm}|p{5mm}|p{5mm}|p{6mm}|p{5mm}|p{5mm}|p{6mm}}
\hline
\multicolumn{2}{c|}{\multirow{2}{*}{preference(\%)$\uparrow$}} &\multicolumn{3}{c|}{Insert} &\multicolumn{3}{c}{Replace}\\
\cline{3-8}
\multicolumn{2}{c|}{}& Base. & Prop. & Equal & Base. & Prop. & Equal \\
\hline
\multirow{4}{*}{VCTK} &Base.1 &27.1&\textbf{53.8}&19.1 &30.2&\textbf{47.6}&22.2\\
\cline{2-8}
~ &Base.2 &37.1&\textbf{47.3}&15.6 &31.8&\textbf{48.7}&19.6\\
\cline{2-8}
~ &Base.3 &35.6&\textbf{49.8}&14.7 &35.6&\textbf{48.0}&16.4\\
\cline{2-8}
~ &Base.4 &34.0&\textbf{52.7}&13.3 &35.1&\textbf{45.3}&19.6\\
\hline
\multirow{4}{*}{MST} &Base.1 &27.7 &\textbf{62.7} &9.7 &25.3 &\textbf{61.3} &13.3\\
\cline{2-8}
~ &Base.2 &9.5 &\textbf{75.8} &14.7 &8.0 &\textbf{81.0} &11.0\\
\cline{2-8}
~ &Base.3 &28.5 &\textbf{50.8} &20.7 &18.7 &\textbf{61.7} &19.7\\
\cline{2-8}
~ &Base.4 &\textbf{39.7} &29.7 &30.7 &26.2 &36.2 &\textbf{37.7}\\
\hline
\end{tabular}
\vspace{-1em}
\end{small}
\end{table}

For subjective evaluation, we test the insertion and replacement operations with VCTK (English) and MST (Chinese) datasets. For each dataset, 15 samples are provided for each operation respectively. Each samples contain 6 audios: the original audio, the edited audio from the baseline system 1 to 4 and proposed system. 15 and 20 listeners participate in the test of VCTK and MST datasets respectively. For each sample, users are required to: 1) mark the naturalness of the edited speech. The mark is from 1 to 5, where 1 means ``completely unnatural'' and 5 means ``completely natural''. 2) mark the similarity of edited audio to the original audio in the unmodified region. The mark is from 1 to 5, where 1 means ``completely different in the unmodified region'' and 5 means ``exactly same in the unmodified region''. 3) indicate the preference between baseline systems and proposed system. The results are shown in Table \ref{tab:insert_replace_text_MOS} and \ref{tab:insert_repalce_text_preference}.

The results show that, EditSpeech achieves the highest MOS in most cases for both insertion and replacement operations in both Chinese and English datasets, except in two cases the highest MOS is achieved by baseline system 4. Moreover, the ABX preference test indicates that listeners prefer EditSpeech in most cases except in one case the baseline system 4 is preferred. 

EditSpeech system completely outperforms the baseline system 1 to 3, which demonstrates that taking both text and speech context into consideration helps to improve the speech editing performance. Specifically, system 1 generates speech based on the whole text but totally neglects the original speech context. The system 2 and 3 both maintain the original speech context by concatenation, but the speech generation is not conditioned on the speech context. The text context is considered in the speech generation in system 3 but not in system 2. In contrast, our system generates speech based on both text context and speech context. Moreover, our system shows advantage compared to baseline system 4 in most cases, indicating the the consideration of both the left and right context further improves the speech editing performance.

\vspace{-0.5em}
\section{Conclusion}
\vspace{-0.5em}
The EditSpeech system allows the users to perform deletion, insertion and replacement of words in a given speech audio. The use of NTTS approach leads to an effective system design and facilitates unrestricted change of speech content. Partial inference and bidirectional fusion introduces low distortion and maintains speech naturalness. Having demonstrated effectiveness on English and Chinese, the design of EditSpeech is expected to be applicable to many other languages.

\vspace{-0.5em}
\section{Acknowledgements}
\vspace{-0.5em}
This research is partially supported by a Tier 3 funding from ITSP (Ref: ITS/309/18) of the Hong Kong SAR Government, and a Knowledge Transfer Project Fund (Ref: KPF20QEP26) from the Chinese University of Hong Kong.

\vfill\pagebreak

\bibliographystyle{IEEEbib}
\bibliography{mybib}

\begin{thebibliography}{10}

\bibitem{whittaker2004semantic}
Steve Whittaker and Brian Amento,
\newblock ``Semantic speech editing,''
\newblock in {\em Proceedings of the SIGCHI conference on Human factors in
  computing systems}, 2004, pp. 527--534.

\bibitem{rubin2013content}
Steve Rubin, Floraine Berthouzoz, Gautham~J Mysore, Wilmot Li, and Maneesh
  Agrawala,
\newblock ``Content-based tools for editing audio stories,''
\newblock in {\em Proceedings of the 26th annual ACM symposium on User
  interface software and technology}, 2013, pp. 113--122.

\bibitem{baume2018contextual}
Chris Baume, Mark~D Plumbley, Janko {\'C}ali{\'c}, and David Frohlich,
\newblock ``A contextual study of semantic speech editing in radio
  production,''
\newblock {\em International Journal of Human-Computer Studies}, vol. 115, pp.
  67--80, 2018.

\bibitem{jin2017voco}
Zeyu Jin, Gautham~J Mysore, Stephen Diverdi, Jingwan Lu, and Adam Finkelstein,
\newblock ``Voco: Text-based insertion and replacement in audio narration,''
\newblock {\em ACM Transactions on Graphics (TOG)}, vol. 36, no. 4, pp. 1--13,
  2017.

\bibitem{morrison2021context}
Max Morrison, Lucas Rencker, Zeyu Jin, Nicholas~J Bryan, Juan-Pablo Caceres,
  and Bryan Pardo,
\newblock ``Context-aware prosody correction for text-based speech editing,''
\newblock {\em arXiv preprint arXiv:2102.08328}, 2021.

\bibitem{descript}
``Descript,'' https://www.descript.com/, 2020.

\bibitem{moulines1990pitch}
Eric Moulines and Francis Charpentier,
\newblock ``Pitch-synchronous waveform processing techniques for text-to-speech
  synthesis using diphones,''
\newblock {\em Speech communication}, vol. 9, no. 5-6, pp. 453--467, 1990.

\bibitem{su2020hifi}
Jiaqi Su, Zeyu Jin, and Adam Finkelstein,
\newblock ``Hifi-gan: High-fidelity denoising and dereverberation based on
  speech deep features in adversarial networks,''
\newblock {\em arXiv preprint arXiv:2006.05694}, 2020.

\bibitem{shen2018natural}
Jonathan Shen, Ruoming Pang, Ron~J Weiss, Mike Schuster, Navdeep Jaitly,
  Zongheng Yang, Zhifeng Chen, Yu~Zhang, Yuxuan Wang, Rj~Skerrv-Ryan, et~al.,
\newblock ``Natural tts synthesis by conditioning wavenet on mel spectrogram
  predictions,''
\newblock in {\em 2018 IEEE International Conference on Acoustics, Speech and
  Signal Processing (ICASSP)}. IEEE, 2018, pp. 4779--4783.

\bibitem{ping2018deep}
Wei Ping, Kainan Peng, Andrew Gibiansky, Sercan~O Arik, Ajay Kannan, Sharan
  Narang, Jonathan Raiman, and John Miller,
\newblock ``Deep voice 3: 2000-speaker neural text-to-speech,''
\newblock {\em Proc. ICLR}, pp. 214--217, 2018.

\bibitem{yu2019durian}
Chengzhu Yu, Heng Lu, Na~Hu, Meng Yu, Chao Weng, Kun Xu, Peng Liu, Deyi Tuo,
  Shiyin Kang, Guangzhi Lei, et~al.,
\newblock ``Durian: Duration informed attention network for multimodal
  synthesis,''
\newblock {\em arXiv preprint arXiv:1909.01700}, 2019.

\bibitem{ren2020fastspeech}
Yi~Ren, Chenxu Hu, Tao Qin, Sheng Zhao, Zhou Zhao, and Tie-Yan Liu,
\newblock ``Fastspeech 2: Fast and high-quality end-to-end text-to-speech,''
\newblock {\em arXiv preprint arXiv:2006.04558}, 2020.

\bibitem{li2019neural}
Naihan Li, Shujie Liu, Yanqing Liu, Sheng Zhao, and Ming Liu,
\newblock ``Neural speech synthesis with transformer network,''
\newblock in {\em Proceedings of the AAAI Conference on Artificial
  Intelligence}, 2019, vol.~33, pp. 6706--6713.

\bibitem{kim2020glow}
Jaehyeon Kim, Sungwon Kim, Jungil Kong, and Sungroh Yoon,
\newblock ``Glow-tts: A generative flow for text-to-speech via monotonic
  alignment search,''
\newblock {\em Advances in Neural Information Processing Systems}, vol. 33,
  2020.

\bibitem{wang2021patnet}
Shiming Wang, Zhenhua Ling, Ruibo Fu, Jiangyan Yi, and Jianhua Tao,
\newblock ``Patnet: A phoneme-level autoregressive transformer network for
  speech synthesis,''
\newblock in {\em ICASSP 2021-2021 IEEE International Conference on Acoustics,
  Speech and Signal Processing (ICASSP)}. IEEE, 2021, pp. 5684--5688.

\bibitem{zheng2019forward}
Yibin Zheng, Jianhua Tao, Zhengqi Wen, and Jiangyan Yi,
\newblock ``Forward--backward decoding sequence for regularizing end-to-end
  tts,''
\newblock {\em IEEE/ACM Transactions on Audio, Speech, and Language
  Processing}, vol. 27, no. 12, pp. 2067--2079, 2019.

\bibitem{zhou2019synchronous}
Long Zhou, Jiajun Zhang, and Chengqing Zong,
\newblock ``Synchronous bidirectional neural machine translation,''
\newblock {\em Transactions of the Association for Computational Linguistics},
  vol. 7, pp. 91--105, 2019.

\bibitem{valin2019lpcnet}
Jean-Marc Valin and Jan Skoglund,
\newblock ``Lpcnet: Improving neural speech synthesis through linear
  prediction,''
\newblock in {\em ICASSP 2019-2019 IEEE International Conference on Acoustics,
  Speech and Signal Processing (ICASSP)}. IEEE, 2019, pp. 5891--5895.

\bibitem{prenger2019waveglow}
Ryan Prenger, Rafael Valle, and Bryan Catanzaro,
\newblock ``Waveglow: A flow-based generative network for speech synthesis,''
\newblock in {\em ICASSP 2019-2019 IEEE International Conference on Acoustics,
  Speech and Signal Processing (ICASSP)}. IEEE, 2019, pp. 3617--3621.

\bibitem{yamamoto2020parallel}
Ryuichi Yamamoto, Eunwoo Song, and Jae-Min Kim,
\newblock ``Parallel wavegan: A fast waveform generation model based on
  generative adversarial networks with multi-resolution spectrogram,''
\newblock in {\em ICASSP 2020-2020 IEEE International Conference on Acoustics,
  Speech and Signal Processing (ICASSP)}. IEEE, 2020, pp. 6199--6203.

\bibitem{kumar2019melgan}
Kundan Kumar, Rithesh Kumar, Thibault de~Boissiere, Lucas Gestin, Wei~Zhen
  Teoh, Jose Sotelo, Alexandre de~Br{\'e}bisson, Yoshua Bengio, and Aaron
  Courville,
\newblock ``Melgan: Generative adversarial networks for conditional waveform
  synthesis,''
\newblock {\em arXiv preprint arXiv:1910.06711}, 2019.

\bibitem{kong2020hifi}
Jungil Kong, Jaehyeon Kim, and Jaekyoung Bae,
\newblock ``Hifi-gan: Generative adversarial networks for efficient and high
  fidelity speech synthesis,''
\newblock {\em arXiv preprint arXiv:2010.05646}, 2020.

\bibitem{ljspeech17}
Keith Ito and Linda Johnson,
\newblock ``The lj speech dataset,'' https://keithito.com/LJ-Speech-Dataset/,
  2017.

\bibitem{yamagishi2019cstr}
Junichi Yamagishi, Christophe Veaux, Kirsten MacDonald, et~al.,
\newblock ``Cstr vctk corpus: English multi-speaker corpus for cstr voice
  cloning toolkit (version 0.92),''
\newblock 2019.

\bibitem{shi2020aishell}
Yao Shi, Hui Bu, Xin Xu, Shaoji Zhang, and Ming Li,
\newblock ``Aishell-3: A multi-speaker mandarin tts corpus and the baselines,''
\newblock {\em arXiv preprint arXiv:2010.11567}, 2020.

\bibitem{mcauliffe2017montreal}
Michael McAuliffe, Michaela Socolof, Sarah Mihuc, Michael Wagner, and Morgan
  Sonderegger,
\newblock ``Montreal forced aligner: Trainable text-speech alignment using
  kaldi.,''
\newblock in {\em Interspeech}, 2017, vol. 2017, pp. 498--502.

\bibitem{g2pE2019}
Kyubyong Park and Jongseok Kim,
\newblock ``g2pe,'' https://github.com/Kyubyong/g2p, 2019.

\bibitem{kingma2014adam}
Diederik~P Kingma and Jimmy Ba,
\newblock ``Adam: A method for stochastic optimization,''
\newblock {\em arXiv preprint arXiv:1412.6980}, 2014.

\bibitem{muller2007dynamic}
Meinard M{\"u}ller,
\newblock ``Dynamic time warping,''
\newblock {\em Information retrieval for music and motion}, pp. 69--84, 2007.

\end{thebibliography}

\end{document}